\documentclass[aps,prl,twocolumn,showpacs]{revtex4}
\usepackage{amsfonts}
\usepackage{amsmath}
\usepackage{amsthm}
\usepackage{graphics}
\usepackage{graphicx}
\usepackage{subfigure}
\usepackage{amssymb}
\usepackage{graphics}
\usepackage{graphicx}
\usepackage{color}
\usepackage{subfigure}

\def\kmax{k_{\rm max}}
\newcommand\bet{{g}}
 
\newcommand\alps{{\frac{\hbar^2}{2m}}}

\newcommand\dertt[1]{ \frac{\partial{ #1}}{\partial t} }
\newcommand\gd{\mbox{${\bf \nabla}^{2}$}}
\newcommand\grad{\mbox{${\bf \nabla}$}}
\newcommand\psib{\overline{\psi}}

\begin{document}
\title{Dispersive bottleneck delaying thermalization of turbulent Bose-Einstein Condensates}
\author{Giorgio Krstulovic}
\affiliation{Laboratoire de Physique Statistique de l'Ecole Normale 
Sup{\'e}rieure, \\
associ{\'e} au CNRS et aux Universit{\'e}s Paris VI et VII,
24 Rue Lhomond, 75231 Paris, France}
\author{Marc Brachet}
\affiliation{Laboratoire de Physique Statistique de l'Ecole Normale 
Sup{\'e}rieure, \\
associ{\'e} au CNRS et aux Universit{\'e}s Paris VI et VII,
24 Rue Lhomond, 75231 Paris, France}
\date{\today}
\begin{abstract}
A new mechanism of thermalization involving a direct energy cascade is obtained in the truncated Gross-Pitaevskii dynamics.
A long transient with partial thermalization at small-scales is observed before the system reaches equilibrium.
Vortices are found to disappear as a prelude to final thermalization. 
A bottleneck that produces spontaneous effective self-truncation and delays thermalization is characterized when large dispersive effects are present at the truncation wavenumber. 
Order of magnitude estimates indicate that self-truncation takes place in turbulent Bose-Einstein condensates.
This effect should also be present in classical hydrodynamics and models of turbulence.
\end{abstract}

\maketitle
The Gross-Pitaevskii equation (GPE) furnishes a dynamical description of superfluids and Bose-Einstein Condensates (BEC) that is valid at very low temperatures \cite{Proukakis:2008p1821}. 
The GPE dynamics is known to produce an energy cascade that leads to a Kolmogorov regime of turbulence \cite{Nore:1997p1333&Nore:1997p1331,Kobayashi:2005p4037}. Such turbulent regimes were studied in low-temperature experiments in superfluid $^4{\rm He}$
\cite{Maurer:1998p4365} and in BEC \cite{Henn:2009p5720}.
The truncated GPE (TGPE), obtained by performing a truncation of Fourier modes, can also describe the (classical) thermodynamic equilibrium of homogeneous BEC \cite{Davis:2001p1475}. The TGPE (microcanonical) equilibrium is known to present (when varying the energy) a condensation transition \cite{Davis:2001p1475,Connaughton:2005p1744&During:2009PhysicaD}.

In the context of classical fluids, the (conservative) truncated Euler dynamics is known to possess long-lasting transients describing dissipative phenomena \cite{Cichowlas:2005p1852}. 
With this motivation, we study in this Letter
the TGPE thermalization that arises from the GPE turbulent energy cascade.
Here is a short summary of our results. Partial thermalization is observed at small-scales during a long transient regime; vortex lines then disappear and final thermalization sets in. A bottleneck that delays the final thermalization is characterized when large dispersive effects are present at truncation wavenumber.

The TGPE is obtained from the GPE describing a homogeneous BEC of volume $V$ by truncating the Fourier transform of the wavefunction $\psi$: $\hat{\psi}_{\bf k}\equiv0$ for $|{\bf k}|>\kmax$ \cite{Davis:2001p1475,Proukakis:2008p1821}.
Introducing the Galerkin projector $\mathcal{P}_{\rm G}$ that reads in Fourier space $\mathcal{P}_{\rm G} [ \hat{\psi}_{\bf k}]=\theta(\kmax-|{\bf k}|)\hat{\psi}_{\bf k}$ with $\theta(\cdot)$ the Heaviside function, the TGPE explicitly reads
\begin{equation}
i\hbar\dertt{\psi} =\mathcal{P}_{\rm G} [- \alps \gd \psi + \bet\mathcal{P}_{\rm G} [|\psi|^2]\psi ],
\label{Eq:TGPEphys}
\end{equation}
where $m$ is the mass of the condensed particles and $g=4 \pi \tilde{a} \hbar^2 / m$, with $\tilde{a}$ the $s$-wave scattering length. 
Madelung's transformation $\psi({\bf x},t)=\sqrt{\frac{\rho({\bf x},t)}{m}}\exp{[i \frac{m}{\hbar}\phi({\bf x},t)]}$ relates the (complex) wavefunction $\psi$ to a superfluid of density $\rho({\bf x},t)$ and velocity ${\bf v}={\bf \nabla} \phi$, where $h/m$ is the Onsager-Feynman quantum of velocity circulation around the $\psi=0$ vortex lines \cite{Proukakis:2008p1821}.
When Eq.\eqref{Eq:TGPEphys} is linearized around a constant $\psi= \hat{\psi}_{\bf 0}$, the sound velocity is given by $c={(g| \hat{\psi}_{\bf 0}|^2/m)}^{1/2}$ with dispersive effects taking place at length scales smaller than the coherence length $\xi={(\hbar^2/2m|\hat{\psi}_{\bf 0}|^2g) }^{1/2}$ that also corresponds to the vortex core size.
In the TGPE numerical simulations presented in this Letter the density $\rho=m N/V$ is fixed to $1$ and the physical constants in eqs.\eqref{Eq:TGPEphys} are determined by the values of $\xi \kmax$ and $c=2$. The quantum of circulation $h/m$ has the value $c\,\xi/\sqrt{2}$ and $V=(2\pi)^3$.

Equation (\ref{Eq:TGPEphys}) exactly conserves the energy $H=\int d^3 x\left( \alps |\grad \psi |^2 +\frac{g}{2}[\mathcal{P}_{\rm G}|\psi|^2]^2 \right)$ and the number of particles $N=\int d^3 x|\psi|^2$. Using Fourier pseudo-spectral methods the momentum ${\bf P}=\frac{i\hbar}{2}\int d^3x\left( \psi {\bf \nabla}\psib - \psib {\bf \nabla}\psi\right)$ is also conserved with dealiasing performed by the $2/3$-rule ($\kmax=2/3\times M/2$ \cite{Got-Ors} at resolution $M$). 

We now study the thermalization of the superfluid Taylor-Green (TG) vortex,
a flow which develops from a spatially-symmetric initial condition prepared by a minimization procedure. The TGPE integrations are performed with a dedicated pseudo-spectral code that uses the symmetries to speed up computations (see reference \cite{Nore:1997p1333&Nore:1997p1331}). 
Up to $512^3$ collocation points are used and the coherence length is determined by $\xi \kmax=1.48$.
\begin{figure}[htbp]
\begin{center}
\includegraphics[height=9.5cm]{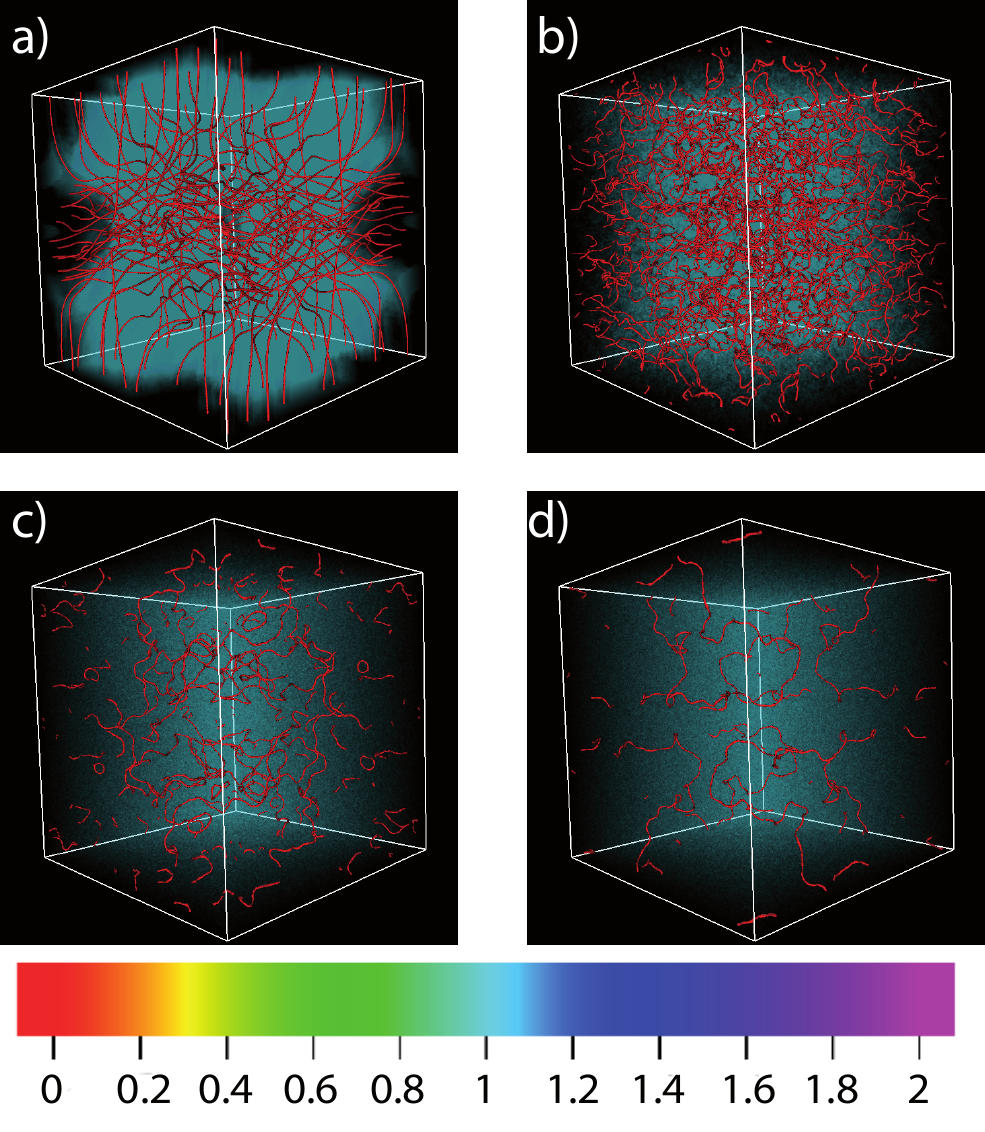}
\caption{$3D$ visualization of density at $t=5$, $10$, $31$ and $55$ at resolution $512^3$ (see below Fig.\ref{Fig:TG}.b). Vortices are displayed as red isosurfaces and clouds correspond to density fluctuations.}
\label{Fig:TG_Phys}
\end{center}
\end{figure}

Vortices and density fluctuations are visualized on Fig. \ref{Fig:TG_Phys}. The short time behavior, see Fig. \ref{Fig:TG_Phys}.a-b, corresponds to the GPE superfluid turbulent regime previously studied in \cite{Nore:1997p1333&Nore:1997p1331}. A new TGPE thermalization regime where vortices first reconnect into simpler structures and then decrease in size with the emergence of a thermal cloud is present at latter times, see Fig.\ref{Fig:TG_Phys}.c-d.

\begin{figure}[htbp]
\begin{center}
\includegraphics[height=6.8cm]{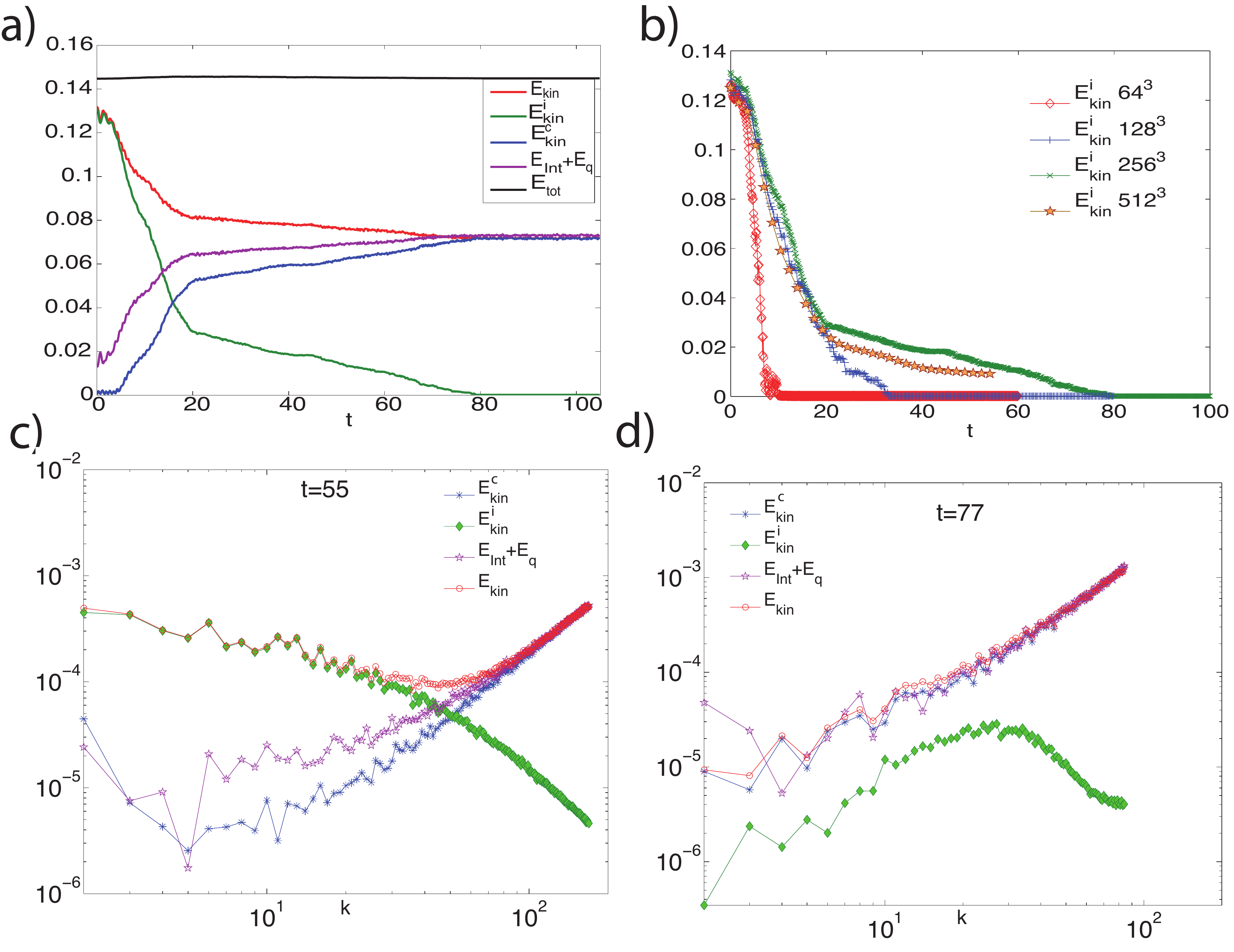}
\caption{a) Temporal evolution of energies $E_{\rm kin}^{\rm c}$,  $E_{\rm kin}^{\rm i}$,  $E_{\rm kin}$ and  $E_{\rm q}+E_{\rm int}$.  At large times, the incompressible energy vanishes and equipartition of energy between $E_{\rm kin}$ and  $E_{\rm q}+E_{\rm int}$ is observed. Resolution of $256^3$. b) Temporal evolution of $E_{\rm kin}^{\rm i}$ at resolution of $64^3$, $128^3$, $256^3$ and $512^3$ with constant $\xi \kmax=1.48$. c-d) Energy spectra at $t=55$ and $t=77$ resolution $512^3$ and $256^3$ respectivelly.}
\label{Fig:TG}
\end{center}
\end{figure}

To further study this relaxation dynamics, we express the energy per unit volume $E_{\rm tot}=(H-mc^2N)/V+\frac{m c^2}{2}$ as the sum of three (space-averaged) parts \cite{Nore:1997p1333&Nore:1997p1331}:
the kinetic energy
$E_{\rm kin} = \langle1/2 (\sqrt \rho v_j)^2\rangle$, the internal energy 
$E_{\rm int}=\langle(c^2/2) (\rho -1)^2\rangle$
and the quantum energy $E_{\rm q}= \langle c^2 \xi^2 (\partial_j \sqrt{\rho} )^2\rangle$.
Parseval's theorem allows to define corresponding energy spectra:
\emph{e.g.} the kinetic energy spectrum
$E_{\rm kin}(k)$ as the (solid angle integral) of
$\left|\frac{1}{2(2\pi)^3} \int d^3 r e^{i r_j k_j}\sqrt \rho v_j \right|^2$.
$E_{\rm kin}(k)$ can be further decomposed into compressible $E_{\rm kin}^{\rm c}(k)$ and incompressible $E_{\rm kin}^{\rm i}(k)$	parts, using
$(\sqrt \rho v_j)=(\sqrt \rho v_j)^{\rm c}+(\sqrt \rho v_j)^{\rm i}$ with $\nabla \cdot (\sqrt \rho v_j)^{\rm i}=0$.
The temporal evolution of $E_{\rm kin}$, $E_{\rm kin}^{\rm i}$, $E_{\rm kin}^{\rm c}$, $E_{\rm q}+E_{\rm int}$ is displayed in Fig.\ref{Fig:TG}.a and the corresponding energy spectra on Fig.\ref{Fig:TG}.c-d.

Three evolution phases are apparent on Fig.\ref{Fig:TG}.a. 
The first phase, for $t\lesssim15$, corresponds to the GPE regime previously studied in \cite{Nore:1997p1333&Nore:1997p1331}.
In the second phase, for $20\lesssim t\lesssim 70$, the spectral convergence of the GP partial differential equation is lost and the dynamics is influenced by $\kmax$. Partial thermalization starts to take place at large wavenumbers where $E_{\rm kin}(k)\sim k^2$ (see Fig.\ref{Fig:TG}.c). Figure \ref{Fig:TG}.b shows that this phase is delayed when the resolution is increased at constant $\xi\kmax$.
When $t>80$ the system reaches the thermodynamic equilibrium with equipartition of energy between $E_{\rm kin}^{\rm c}$ and $E_{\rm q}+E_{\rm int}$, see Fig. \ref{Fig:TG}.d. Finally, $E_{\rm kin}^{\rm i}$ vanishes before final thermalization (see Fig.\ref{Fig:TG}a-b); the total disappearance of vortices is observed on corresponding $3D$ visualizations (data not shown). 
Similar relaxation mechanisms are also present in models of hydrodynamic turbulence and the truncated Euler dynamics \cite{Connaughton:2004p4597,Frisch:2008p1877,Cichowlas:2005p1852}.

\begin{figure}[htbp]
\begin{center}
\includegraphics[height=5cm]{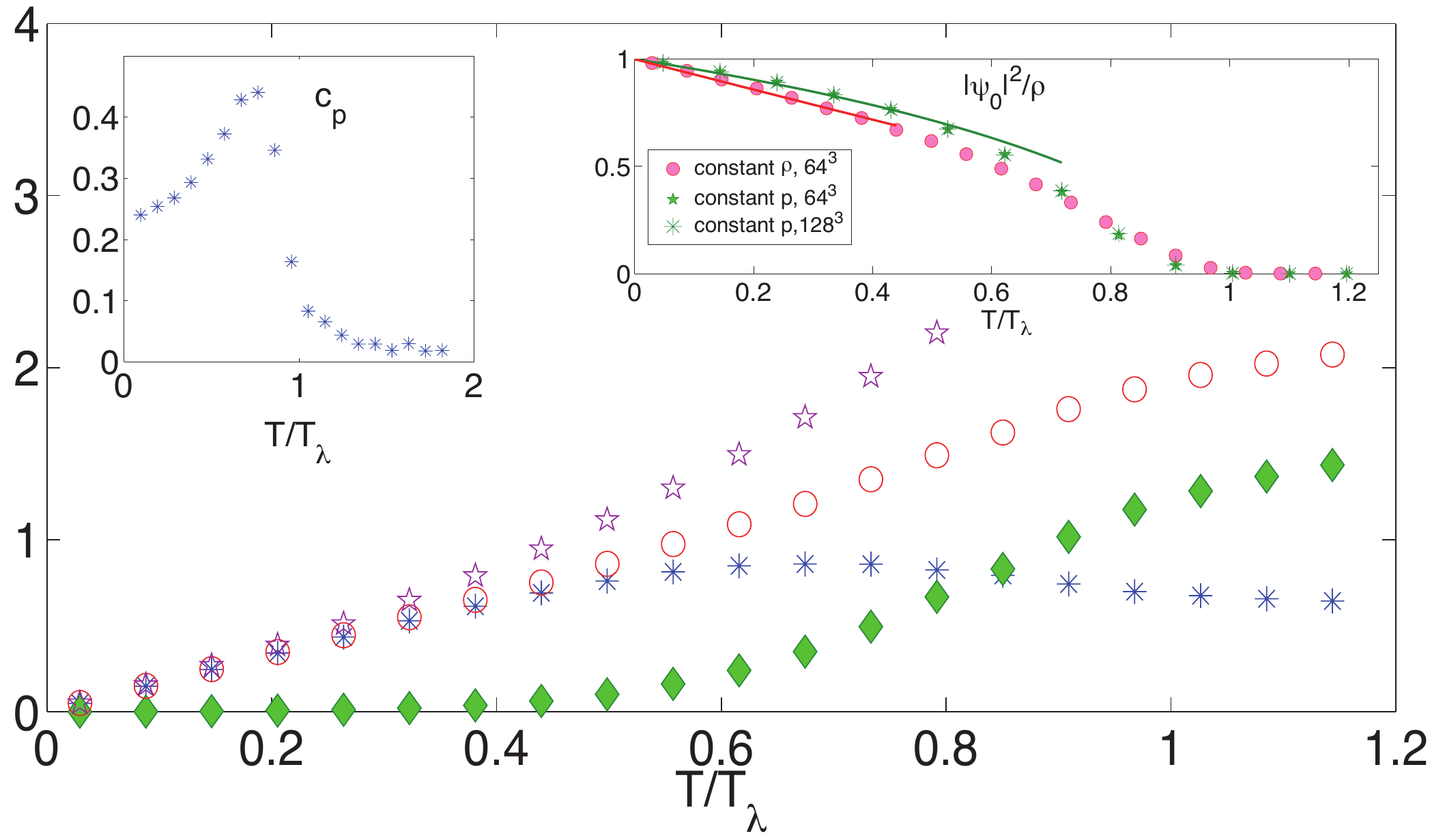}
\caption{Temperature dependence of the energies $E_{\rm kin}^{\rm c}$ (stars),  $E_{\rm kin}^{\rm i}$ (diamonds),  $E_{\rm kin}$ (circles) and  $E_{\rm q}+E_{\rm int}$ (pentagrams) at constant density. Insets: (right) Temperature dependence of the condensate fraction $|\psi_{\bf 0}|^2/\rho$; (left) Specific heat $c_{p}=\left.\frac{\partial H}{\partial T}\right|_{p}$ at resolution $128^3$.}
\label{Fig:Scan}
\end{center}
\end{figure}

We now focus on a characterization of thermodynamic equilibrium that will allow us to account for the absence of vortices and the equipartition of energy in the final states.
Microcanonical equilibrium states are known to result from long-time integration of TGPE \cite{Davis:2001p1475,Connaughton:2005p1744&During:2009PhysicaD}. Grand canonical equilibrium states are given by the probability distribution $\mathbb{P}_{\rm st}[\psi]=\mathcal{Z}^{-1}\exp[{-\beta (H-\mu N)}]$. They allow to directly control the temperature (instead of the energy in a microcanonical framework). These states can be efficiently obtained by constructing a stochastic process that converges to a realization with the probability $\mathbb{P}_{\rm st}[\psi]$ \cite{Krstulovic:PhdThesis}. This process is defined by a Langevin equation consisting in a stochastic Ginbzurg-Landau equation (SGLE):
\begin{eqnarray}
\nonumber\hbar\dertt{\psi} &=&\mathcal{P}_{\rm G} \left[\alps \gd \psi+\mu \psi - \bet\mathcal{P}_{\rm G} [|\psi|^2] \psi\right] \\
&&+\sqrt{\frac{2 \hbar}{V\beta }} \mathcal{P}_{\rm G} \left[\zeta({\bf x},t)\right] \label{Eq:SGLRphys},\hspace{3mm}
\end{eqnarray}
where the white noise $\zeta({\bf x},t)$ satisfies $\langle\zeta({\bf x},t)\zeta^*({\bf x'},t')\rangle=\delta(t-t') \delta({\bf x}-{\bf x'})$, $\beta$ is 
the inverse temperature and $\mu$ the chemical potential.
Using this algorithm in \cite{Krstulovic:PhdThesis} the microcanonical and grand canonical ensembles were shown to be equivalent and the condensation transition reported in \cite{Davis:2001p1475,Connaughton:2005p1744&During:2009PhysicaD} identified with the standard second order $\lambda$-transition (see insets on Fig.\ref{Fig:Scan}). Note that $c_{p}$ would be very difficult to obtain from microcanonical results. 

At low-temperature the partition function $\mathcal{Z}$ can be exactly computed by the steepest-descent method \cite{Krstulovic:PhdThesis}. In particular, the mean value of the condensate amplitude reads 
$\overline{|\hat{\psi}_{\bf 0}|^2}=\frac{\mu}{g}-\frac{\mathcal{N}}{V\beta \mu}f_{0}\left[\frac{4 m \mu
  }{\hbar^2k_{\rm max}^2}\right]$, 
where $\mathcal{N}=k_{\rm max}^3 V/6 \pi ^2$ is the total number of modes and $f_0[z]=3 z- 9 z^{3/2} \cot ^{-1}\left(\sqrt{z}\right)/4$.

In numerical simulations of Eq.\eqref{Eq:SGLRphys}, $\mu$ is adjusted to fix the density (or the pressure $p$) \cite{Krstulovic:PhdThesis}. The inverse temperature is normalized as $\beta=\mathcal{N}/VT$. With this choice of parametrization the $\lambda$-transition temperature $T_\lambda$ is independent of $\mathcal{N}$. Data from SGLE and low-temperature calculation are confronted on Fig.\ref{Fig:Scan} and seen to be in good agreement.

The temperature dependence of the different energies is displayed on Fig.\ref{Fig:Scan}. Observe that $E_{\rm kin}^{\rm i}$ vanishes at temperatures $T/T_{\lambda}\lesssim1/2$. This explains the disappearance of vortices in Fig.\ref{Fig:TG} above as the corresponding final temperature is well below $T_{\lambda}$ (see corresponding values of energies on Fig.\ref{Fig:TG}.a). At low temperature equipartition of energy between $E_{\rm kin}$ and $E_{\rm q}+E_{\rm int}$ is also apparent on Fig.\ref{Fig:Scan}.
Note that a larger $\kmax$ implies, by equipartition, a lower temperature. The corresponding dissipative effects are thus smaller explaining the thermalization delay apparent on Fig.\ref{Fig:TG}.b.

We now turn to the study of dispersive effects on the thermalization of the TGPE dynamics. 
In order to investigate dispersive effects, the TG initial condition described above is prepared at fixed $\xi=\sqrt{2}/20$ and varying resolution: $64^3$, $256^3$ and $256^3$. 
The three initial condition thus represent the same field at different resolutions.

\begin{figure}[htbp]
\begin{center}
\includegraphics[height=10cm]{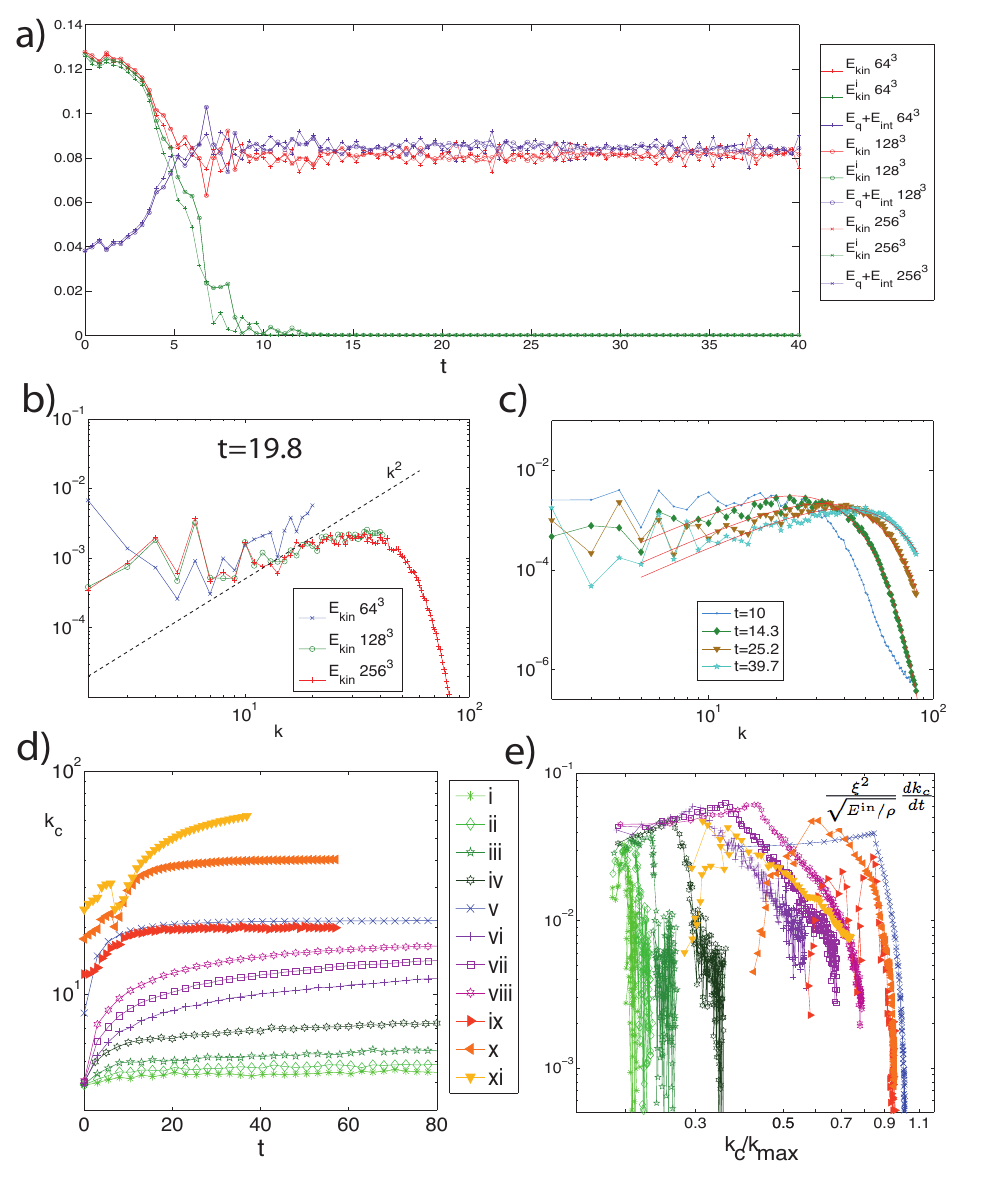}
\caption{a) Evolution of energies at $\xi=\sqrt{2}/20$ and resolution $64^3$, $128^3$ and $256^3$. b) Energy spectrum $E_{\rm kin}(k)$ at $t=19.8$ for the three TG runs. c) Evolution of $E_{\rm kin}(k)$, solid red lines are fits of the form $A k^{2}\exp{[-\gamma^2 k^2]}$ (see text). d) Evolution of $k_c$. Curves  i-iv: $\xi=2\sqrt{2}/5$, $k_c^{\rm in}=4$, $E^{\rm in}=0.1,\,0.2\,,0.4\,,1$; v: $\xi=\sqrt{2}/10$, $k_c^{\rm in}=8$, $E^{\rm in}=0.2$; vi-viii: $\xi=\sqrt{2}/5$,  $E^{\rm in}=0.1,\,0.2\,,0.4$ (i-viii in resolution $64^3$); ix-xi: Taylor-Green resolutions $64^3$, $128^3$ and $256^3$.  e) Parametric representation $dk_{c}/dt$ v.s. $k_{c}/\kmax$ (same labels as d).}
\label{Fig:SelfTrunc}
\end{center}
\end{figure}

The evolutions of the energies of the three runs are shown on Fig.\ref{Fig:SelfTrunc}.a. 
They are identical until $t\approx5$ where the run at resolution $64^3$ starts to lose spectral convergence.
All runs appear to have completely thermalized at $t\approx20$. 
However the kinetic energy spectra corresponding to this time, displayed on Fig.\ref{Fig:SelfTrunc}.b, shows clear differences between runs. The high-wavenumber modes are thermalized in the $64^3$ run but they decay at large-$k$ at higher resolutions. 
In the $256^3$ run, two zones are clearly distinguished: an intermediate thermalized range (with approximative $k^2$ scaling) followed, well before $\kmax=85$, by a steep decay zone. 

The temporal evolution of $E_{\rm kin}(k)$ for the $256^3$ run displayed in Fig.\ref{Fig:SelfTrunc}.c. is well represented by a fit of the form $A(t) k^{2}\exp{[-\gamma^2(t)k^2]}$, where $A(t)$ and $k_c(t) \sim \gamma^{-1}(t)\ll \kmax$ are increasing functions of $t$.
Such a behavior of the energy spectra ensures spectral convergence and the dynamics is thus not influenced by $\kmax$. 
Furthermore, we checked that $A(t) k_c^3(t)\approx E_{\rm kin}$ for $t\gtrsim 20$ (data not shown). 
This new regime can be described as a (quasi) thermalization, with self-truncation at wavenumber $k_c$ and temperature $T\sim E/k_c^3$, that spontaneously establishes itself within the GP partial differential equation dynamics
when the direct energy cascade is inhibited by a dispersive bottleneck for the energy transfer \footnote{A different kind of bottleneck was proposed in \cite{Lvov:2007p3036}, with a inhibition of the energy transfer involving Kelvin wave cascading along vortices rather than dispersive effects.}.

An open question is whether thermalization is simply delayed or, as in the Fermi-Pasta-Ulam-Tsingu problem \cite{Fermi:1955p3871}, completely inhibited in the self-truncation regime $\xi k_{\rm max}\to \infty$.
 It is not feasible now to directly study this limit, within the TG framework, as it requires long runs at arbitrarily high resolution.
To skip the TG cascade regime and directly study the self-truncated thermalization regime we use initial data generated by the SGLE instead of the TG vortex.
To wit, we use Eq.\eqref{Eq:SGLRphys} with a variable truncation wavenumber $k_c^{\rm in}$, set to a target value of $k_c$, smaller than the maximum truncation wavenumber $k_{\rm max}$ allowed by the resolution. 
This SGLE-generated initial data is then used to run the TGPE at a given value of $\xi k_c$ with arbitrarily large values of $\xi k_{\rm max}$. 

A number of such runs were performed at resolution $64^3$ with various values of $k_c^{\rm in}$, $\xi$, and initial energy $E^{\rm in}$. 
The result of these computations are compared with the TG runs and displayed on Fig.\ref{Fig:SelfTrunc}.d. The self-truncation wavenumber is explicitly determined from $E_{\rm kin}$ by the integral formula
$k_c^2=\frac{5}{3}\int_0^{\kmax} k^2E_{\rm kin}(k)dk/\int_0^{\kmax} E_{\rm kin}(k)dk$.
A general growth in time of $k_c$ is apparent on Fig.\ref{Fig:SelfTrunc}.d for both the Taylor-Green runs and the SGLE-generated initial data, showing similar behavior. 

In order to check for a self-similar regime a parametric Log-Log representation 
of $dk_{c}/dt$ v.s. $k_{c}$ is used on Fig.\ref{Fig:SelfTrunc}.e. With this representation, a self-similar evolution $k_c(t)\sim t^\alpha$ corresponds to a line of slope $(\alpha-1)/\alpha$. Figure \ref{Fig:SelfTrunc}.e shows transient self-similar evolutions terminated by a vertical asymptote corresponding to logarithmic growth ($\alpha=0$). This self-truncation takes place at small values of $k_{\rm c}/\kmax$ strongly suggesting that the self-truncation happens in a regime independent of cut-off.
This regime should, in principle, be amenable to a description in terms of wave turbulence theory along the lines of reference \cite{svistunov}.

Let us concentrate now on estimations of order of magnitude relevant to physical weakly-interacting BEC. 
At very low-temperature, the GPE gives an accurate description of the (classical) dynamics of BEC \cite{Proukakis:2008p1821} at scales larger than the mean inter-atomic particle distance $\ell \sim |\hat{\psi_{\bf 0}}|^{-2/3}$, satisfying $\tilde{a}\ll\ell\ll \xi$. 
%
At finite temperature \cite{Davis:2001p1475,Proukakis:2008p1821},
Bogoluibov's dispersion relation
$\omega_{\rm B}^2(k)=k^2 g|\psi_{\bf 0}|^2/m+k^4\hbar^2/4m$
and the 
relation $\hbar\omega_{\rm B}(k_{\rm eq})=\beta^{-1}=k_{\rm B}T$ 
imply that phonons are in equipartition only for wavenumbers $k<k_{\rm eq}$.
The equipartition wavenumber thus satisfies 
$\xi k_{\rm eq} \sim T/T^*$ for $T\ll T^*$ and $\xi k_{\rm eq} \sim \sqrt{T/T^*}$ for $T\gg T^*$,
with $T^*=\ell^2T_\lambda/\xi^2$ and $T_\lambda\sim\hbar^2/k_{\rm B} m\ell^2$ the condensation temperature (see ref. \cite{Proukakis:2008p1821}).
Thus physical BEC at low-temperature have a natural cut-off for the equipartition range given by $\kmax=k_{\rm eq}(T)$.

In experimental BEC the value of $\xi k_{\rm eq}$ is large because $T^*/T_\lambda\sim \ell^2/\xi^2\ll1$ and the corresponding TGPE has a large $\xi\kmax$.
This strongly suggests that, unless overwhelmed by other (non-TGPE) relaxation mechanisms, the thermalization slowdown caused by the dispersive bottleneck should be observable. 
It would thus be interesting to investigate in the future the suppression of aspect-ratio inversion observed in recent turbulent BEC experiments \cite{Henn:2009p5720} in the context of the TGPE in a non-homogenous trapping geometry.

In the context of classical hydrodynamics, self-truncation can take place in fluids with dispersion {\emph e.g.} MHD flows with Alfv\'en waves \cite{Krstulovic:PhdThesis}. 
Models of turbulence of the Euler-$\alpha$ \cite{LinshizTiti} type, where the r.h.s. of the incompressible Euler equation is multiplied by the operator $(1-\alpha \nabla^2)^{-1}$ that penalizes the energy at small-scales, are also plausible candidates for self-truncation.
%

In summary we investigated the thermalization dynamics in turbulent BEC
using the TGPE.
Our main result is that a bottleneck delays the final thermalization when large dispersive effects are present at truncation wavenumber and produces an effective self-truncation. These effects are in principle observable in physical BEC.

We acknowledge useful scientific discussions with G. D\"{u}ring and S. Rica.  
The computations were carried out at IDRIS (CNRS) and visualizations used VAPOR \footnote{http://www.vapor.ucar.edu}.



\end{document}